\documentclass[jkps,preprint,fleqn,showpacs,showkeys]{revtex4}
\usepackage{graphicx}
\usepackage{amssymb}
\usepackage{amsmath}
\usepackage{bm}
\newcommand{\be}{\begin{eqnarray}}
\newcommand{\ee}{\end{eqnarray}}
\begin{document}
\setcounter{page}{0}
\title[]{Phase diagram of the Wako-Sait\^{o}-Mu\~noz-Eaton $\beta$ hairpin Model obtained from the partition function zeros}
\author{Julian \surname{Lee}}
\email{jul@ssu.ac.kr}
\thanks{Fax: +82-2-824-4383}
\affiliation{Department of Bioinformatics and Life Science, Soongsil University, Seoul
156-743}

\date[]{Received 22 July 2014}

\begin{abstract}
I study the partition function zeros of the Wako-Sait\^{o}-Mu\~noz-Eaton (WSME) $\beta$ hairpin model in the complex temperature plane. For various values of the entropy cost of disordering a bond, the zeros show clear locus corresponding to the folding transition. By extrapolating the locus to the real axis, transition temperature can be determined for various values of the entropy cost, leading to the phase diagram of the WSME  $\beta$ hairpin model.
\end{abstract}



\maketitle

\section{INTRODUCTION}

Studying equilibrium and dynamic aspects of protein folding is of great interest in computational biophysics and statistical physics. Due to the computational costs of studying realistic all-atom model, simplified toy models such as lattice proteins have been a subject of extensive research\cite{LD89, BJ04, WL09, CL05}. 

A class of toy models called Go-like models bias the Hamiltonian toward its native structure\cite{WS,M97,M98,M99,A99,G99,B00,F02,BP,D04,I07,I10}. The Go-like models are  based on the assumption that the native structure of a protein contains much information on the folding process. In this work, I study the protein folding transition using  Wako-Sait\^{o}-Mu\~noz-Eaton (WSME) mode\cite{WS,M97,M98,M99}, a simple Go-like model that describes a protein as a one-dimensional Ising-like model with long-range interaction. In order to specify a model, a native structure of a protein of interest should be first given. I will restrict ourselves to a $\beta$-hairpin in this work. 

I apply the partition function zeros method\cite{YL,Fisher,IPZ,Bo,JK,AH,YJK,B03,Wang,CL,BDL,SYK,JL,Ar00,BE03,A90,prd92,prd12, Bor04,Zhu06,PRC08,PRL}, a powerful method for studying a phase transition, to study the folding transition of a $\beta$-hairpin in the context of WSME model. The partition function zeros of WSME $\beta$-hairpin model in the complex temperature plane, called the Fisher zeros\cite{Fisher}, were studied in my previous works\cite{PRL}. The zeros for hairpin with even number of bonds could be obtained analytically in the limit when the entropic cost of ordering a bond is much larger than one\cite{PRL}. The zeros for hairpin with even number of bonds, whose structure has more resemblance to the real $\beta$-hairpins, could be obtained only numerically, and the zeros for the $\beta$-hairpin with 15 bonds were described.

In this work extend the previous results, and obtain the Fisher zeros for a $\beta$-hairpins with 20 and 21 bonds, for various values of entropic costs of ordering a bond. Although the $\beta$-hairpin is a finite-size system, the folding transition temperature can be obtained by extrapolating the locus of zeros to the positive real axis. By plotting the transition temperature for various values of entropic cost, which plays the role of the magnetic field in a magnetic system, I can obtain the phase diagram of WSME $\beta$-hairpin model.

\section{The model}

WSME model describes a protein as a one-dimensional Ising model with long-range interactions\cite{WS,M97,M98,M99}. A binary variable $m_i$ is associated with each bond of a protein which can take either one or zero depending on whether the bond is in the ordered or disordered state. Only the native contact contributes to the energy, and the contact can be formed if and only if all the intervening residues in the sequence are in the ordered states, so the Hamiltonian reads:
\be
{\cal H}(\{m_k\}) = \sum_{i=1}^{N-1} \sum_{j=i+1}^N \epsilon_{i j} \Delta_{i j} \Pi_{k=i}^j m_k 
\ee
where $N$ is the number of bonds, $\epsilon_{i j} < 0$ is the strength of contact between the $i$-th and the $j$-th bond, and $\Delta_{ij}=1$ if the corresponding contact is formed in the native structure and zero otherwise. There is an entropy $\Delta S_i$ difference between an ordered and a disordered bond, and assuming that the ordered bond is described by a unique microstate, $\exp(\Delta S_i)$ the effective number of microstates of a disordered bond. The total number of microstates per bond is then $\Lambda_i\equiv \exp(\Delta S_i) +1$.

The partition function can be written as:
\be
Z = \sum_{E}  \Omega(E; \{ \Lambda_k \} )  \exp(-\beta E)
\ee
where the density of states $\Omega(E; \{ \Lambda_k \} )$ is the number of microstates for given values of the energy $E$ and the parameters $\{ \Lambda_k \}$. Therefore, once the density of states is computed, the partition function is obtained from them in a straightforward manner. 

\begin{figure}
\includegraphics[width=10.0cm]{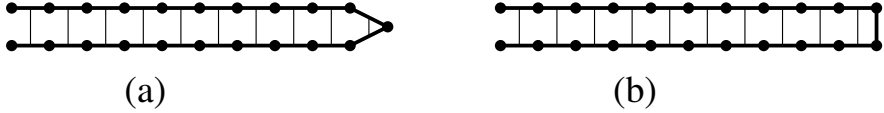}
\caption{The naive conformations of hairpin models with (a) $N=20$ and (b) $N=21$. The residues are denoted as filled circles, and the hydrogen bonds are shown as thin lines.
}\label{hairpin}
\end{figure}

The model is fully determined only after we specify the native structure of protein. In this work I will restrict myself to a simple $\beta$-hairpin with $N=20$ and $N=21$, where $N$ is the number of peptide bonds, and the $i$-th bond makes contact with $N-i+1$-th bond. These hairpins are depicted in Figure \ref{hairpin}.
The conformation with odd value of $N=2n+1$ has the same number of native contacts $n$ as that with $N=2n$ except that it has an extra bond at the turn not participating in the native contact. The conformation with an extra bond at the turn is more realistic, but the model with an even value of $N$ has an advantage that its zeros can be obtained analytically in a certain limit. I will assume that the contacts are formed mainly due to the backbone hydrogen bond and that the sequence effects are negligible, so the interaction strengths are uniform: $\epsilon_{ij} = -\epsilon$. Then the energy of the conformation where $j$ native contacts are broken is simply $E_j = j \epsilon$ where we shifted the energy by a constant amount so that the energy of the fully folded conformation is taken to be zero.   I will also assume that  the entropic cost of ordering each bond is the same throughout the chain: $\Lambda_i = \Lambda$. Then the density of states $\omega_j(\Lambda)$ for a given value $j$ of broken native contacts then takes the form\cite{PRL}:
\be
\omega_j(\Lambda) = \left\{
    \begin{array}{ll}
        1 &  (j=0),\\
        \left(\Lambda^2-1\right)\Lambda^{2j-2} &  (1 \le j \le n\ {\rm for\ even\ }N)\ (1 \le j < n\ {\rm for\ odd\ }N)  \\
        \left(\Lambda^3-1\right)\Lambda^{2n-2} &  (j =n\ {\rm for\ odd\ }N).
    \end{array} \right. \label{DS}
\ee
The partition function is then written as:
\be
Z = \sum_{j}  \omega_j(\Lambda)  y^j \label{bpart}
\ee
where $y \equiv \exp(-\beta \epsilon)$. 

The zeros of the partition function Eq.(\ref{bpart}) in the complex plane of $y$ have been investigated for $N=14$ and $N=15$\cite{PRL}. In this work, I study the zeros of the partition function $Z$ for $N=20$ and $N=21$, with the same value of $n=10$, for various real values of $\Lambda$, to obtain the phase diagram of WSME $\beta$-hairpin model.

\section{The partition function zeros and the phase diagram}
The partition function zeros for a given value of $\Lambda$ are obtained as the ten roots to the polynomial equation:
\be
Z(y) = \sum_{j=0}^{10}  \omega_j(\Lambda)  y^j = 0. \label{pol}
\ee
Since the coefficient are real, any complex roots form conjugate pairs. Note that the summation forms a geometric series for an even value of $N$ if $\omega(0)=1$ is replaced by $\omega(0)=1-1/\Lambda^2$, which can be justified only for $\Lambda >> 1$. In this case Eq.(\ref{pol}) can be solved analytically to yield zeros distributed on a circle~\cite{PRL}. Otherwise, the zeros have to be obtained numerically. It has been shown for $N=14$ that the zeros lies close to the analytic locus even for $\Lambda$ as small as 2.0~\cite{PRL}.
    
\begin{figure}
\includegraphics[width=15.0cm]{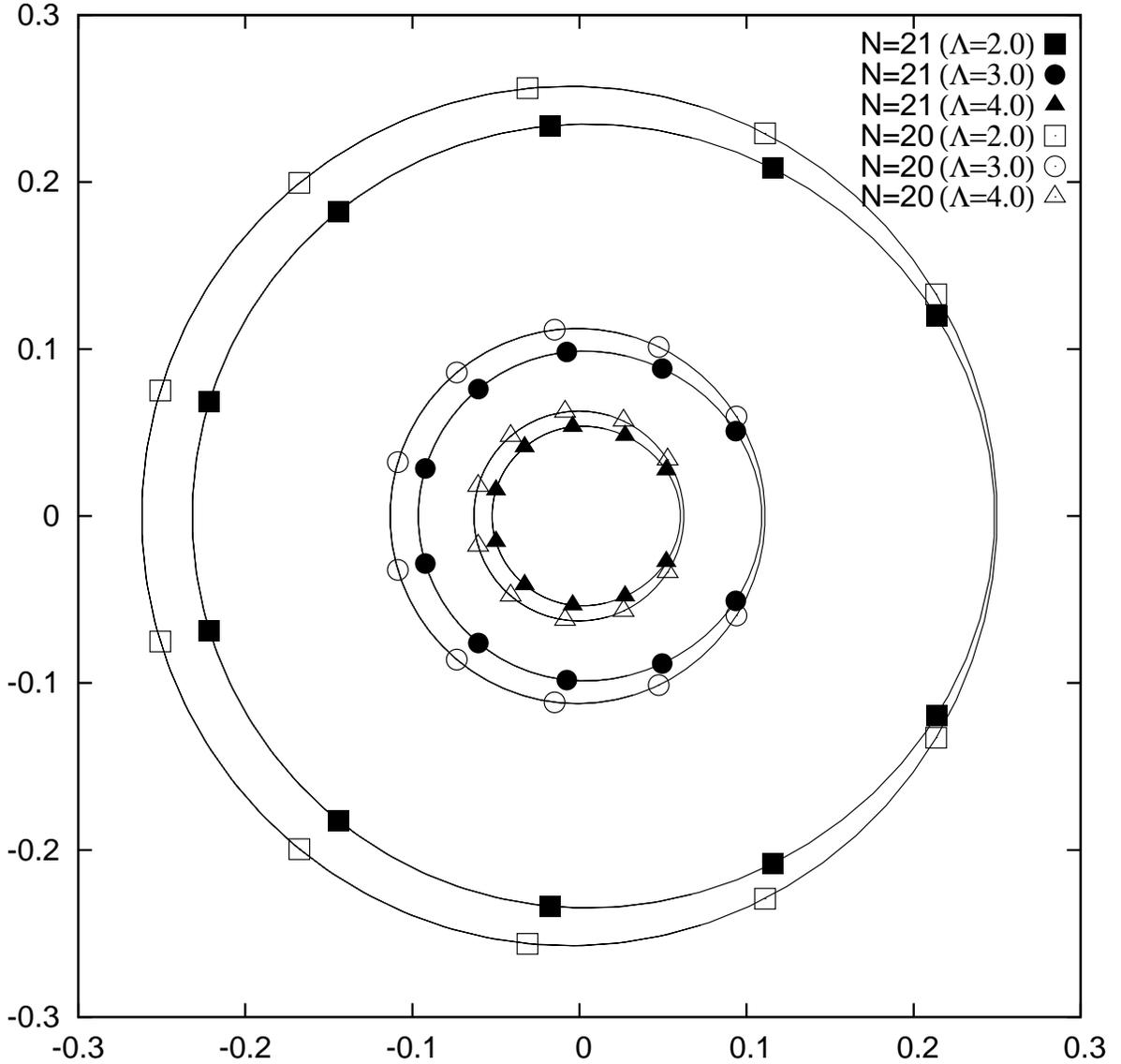}
\caption{The zeros for $N=20$ (open symbols) and $N=21$ (filled symbols). The curves obtained by extrapolating the zeros are also shown.}\label{PFZ}
\end{figure}

The zeros of the WSME $\beta$-hairpin for $N=20$ and $N=21$ are plotted in the Figure \ref{PFZ} for several representative values of $\Lambda$, which were computed using MATHEMATICA. 
The zeros form a clear visual locus that separate the complex plane into two regions. Although the $\beta$-hairpin is a system of finite size, a locus of PFZs can be used to {\it define} the phases of the system\cite{Bo, JK}. Note that the physical range of $y \equiv \exp(-\beta \epsilon)$ is $0 \le y \le 1$, with $y=0$ and $y=1$ corresponding to $T=0$ and $T=\infty$, respectively. Therefore, the region inside and outside the locus define the folded and unfolded phases respectively. 

The loci of zeros closely resembles a circle for $N=20$, but those for $N=21$ have more skewed form, the distance of zero from the origin shorter near the negative real axis compared to those near positive real axis. As to be expected, the folded region shrinks in favor of unfolded region as $\Lambda$ is increased. Since the zeros are exact, I extrapolated the locus to the positive real axis by obtaining a function expanded in terms of trigonometric function:
\be
r(\theta) = \sum_{j=0}^{4} A_j \cos (j \theta) 
\ee
where $r$ and $\theta$ are the polar coordinates in the complex plane of $y$:
\be
y=r e^{i \theta}
\ee
$A_j$s are determined so that the zeros sit on the curve for $r(\theta)$. Since ten zeros form five conjugate pairs, there are only five independent zeros to be used, and only the even function of $\theta$ is allowed due to reflection symmetry with respect to the real axis. The method used here is similar to BST extrapolation\cite{BST,PTVF,Monroe}, except that trigonometric function is used instead of rational function, since we are considering extrapolation with respect to an angle. The curve obtained by extrapolation is shown in Figure \ref{PFZ}.

\begin{figure}
\includegraphics[width=10.0cm]{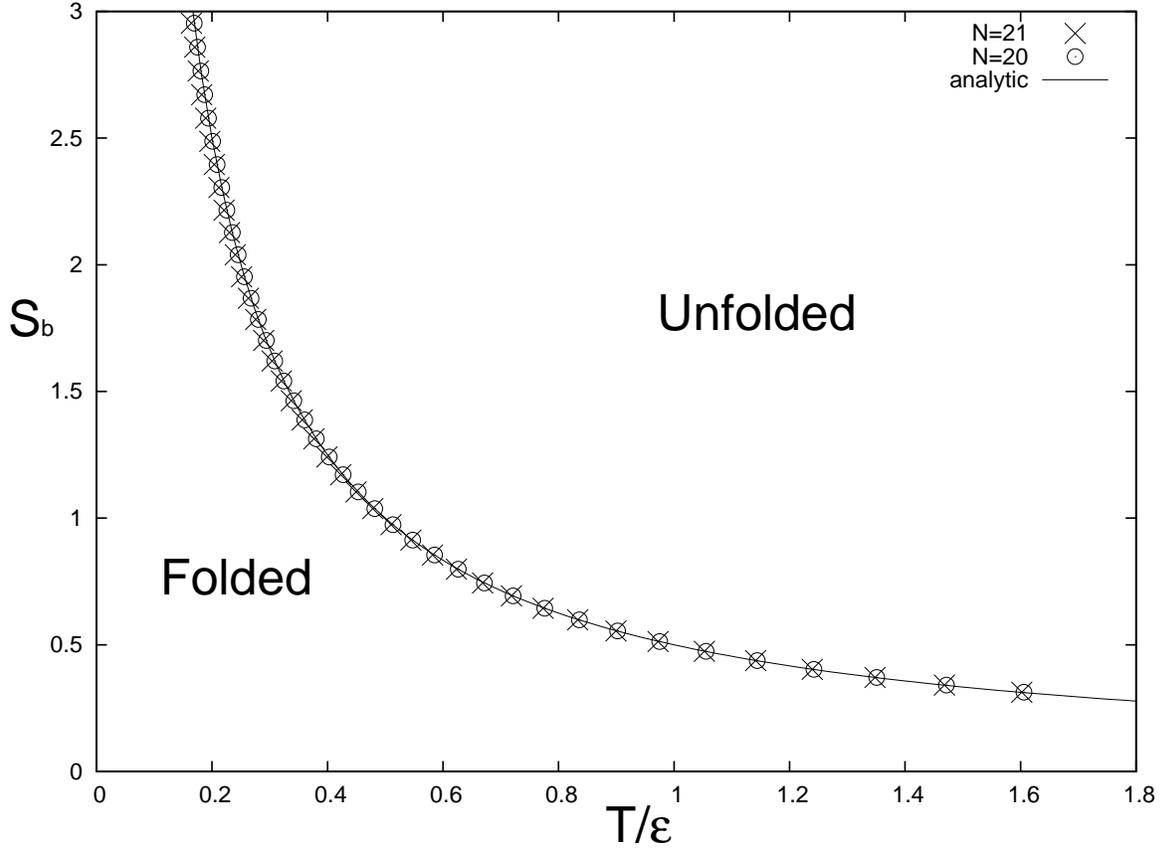}
\caption{The plots of the transition temperature for various values of $S_b=\ln(\Lambda)$, for $N=20$ (circles) and $N=21$ (crosses). The curve obtained by analytic formula  are also shown.}\label{phase}
\end{figure}

The transition temperatures obtained for various values of entropy per bond $S_b \equiv \ln (\Lambda)$ are plotted in Figure \ref{phase}, leading to a phase diagram. The analytic formula of zeros for even values of $N$ have obtained under the assumption of $\Lambda \gg 1$ leads to the folding temperature\cite{PRL}
\be
\frac{T_f}{\epsilon} = \frac{1}{2 S_b}
\ee
which is also shown as a solid line in Figure \ref{phase}.
We see that the phase boundary defined from the exact zeros coincides quite well for $N=20$, even for $s$ as small as $0.31$, corresponding to $\Lambda \simeq 1.37$, although the analytic formula was obtained under the assumption of $\Lambda \gg 1$. Even the phase boundary obtained for $N=21$ coincides very well with the analytic curve, although we see a slight shift toward left for large values of $S_b$. This is to be expected, since the additional entropy cost for the extra bond at the turn will tend to favor the unfolded conformation. However, this effect is quite small, and as can be seen from the loci of zeros (Fig.\ref{PFZ}), the main effect of this extra turn is only to shift the loci near the negative real axis, transition temperature at the positive real axis being quite robust with respect to the existence of the extra bond at the turn.

\begin{figure}
\includegraphics[width=10.0cm]{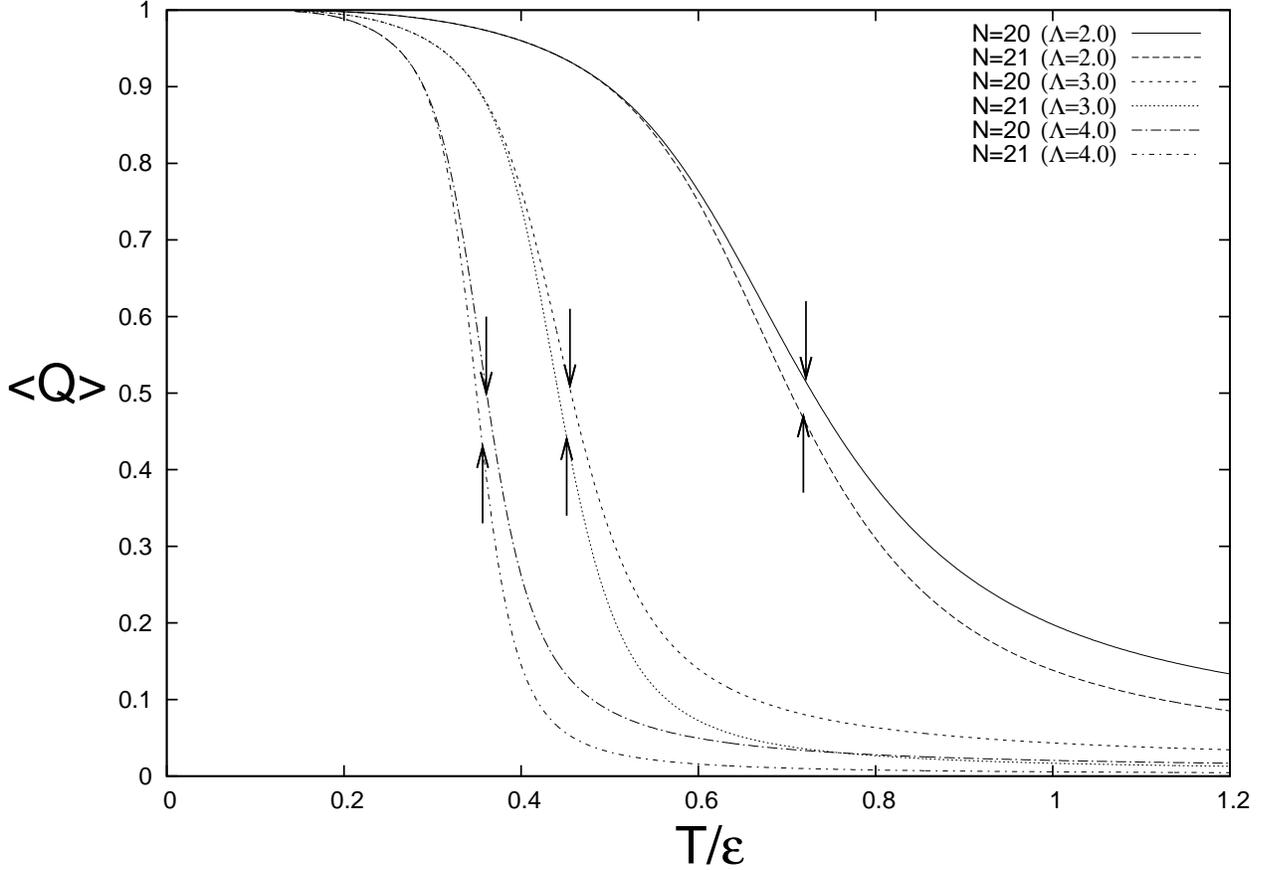}
\caption{The graph of the average fraction of unbroken native contacts  $\langle Q \rangle$ as a function of temperature, for $\Lambda=1,2,3$ and $N=20,21$. The transition temperatures defined by extrapolation of the zeros are indicated by the arrows.}\label{pcont}
\end{figure}

To confirm that the low and high temperature regions correspond to folded and unfolded phases, I also drew the average fraction of unbroken native contacts $\langle Q \rangle$ as functions of temperature (Fig.\ref{pcont}), for representative values of $\Lambda$. The transition temperature defined by the extrapolation of the zeros are shown as arrows, and we see indeed the average fraction of native contacts are about 0.5 at the transition point.

\section{CONCLUSIONS}

In work, I studied the exact partition function of WSME $\beta$-hairpin model for $N=20$ and $N=21$. The model with $N=21$ has the same number of native contacts as that with $N=20$ except that it has an extra bond at the turn not participating in the native contact. I find that for a given value of the parameter $\Lambda$, the effective number of microstates per bond, the locus of the zeros for $N=21$ is skewed near negative real axis compared to that of $N=20$ which is closely approximated as a circle. However, zeros near the positive real axis are quite robust with respect to the existence of the extra bond at the turn. The phase diagram was obtained for various values of $S_0$, the entropy cost for ordering a bond, and indeed the phase boundaries for $N=21$ and $N=20$ agree quite well with each other, as well as the analytic formula for even values of $N$ obtained under the assumption of $\Lambda >> 1$. I could observe a slight shift of the boundary toward the folded region for $N=21$ and large values of $S_0$, as to be expected. I also confirmed that roughly half of the native contacts are formed on average, at the transition temperature obtained by the extrapolation of the partition function zeros. 

\begin{acknowledgments}
This study was supported by a Grant No. A120757 of the Korea Healthcare Technology R\&D, Ministry of Health and Welfare, Republic of Korea.
\end{acknowledgments}

\end{document}